\documentclass[11pt]{article}
\usepackage{verbatim}
\usepackage{amsmath,amssymb}
\makeatletter \@addtoreset{equation}{section} \makeatother

\newcommand{\noi}{\vspace{12pt}\noindent}
\newcommand{\beq}{\begin{equation}}
\newcommand{\eeq}{\end{equation}}
\newcommand{\bea}{\begin{eqnarray}}
\newcommand{\eea}{\end{eqnarray}}

\newcommand{\e}[1]{{(\ref{#1})}}
\newcommand{\eq}[1]{{eq.\ (\ref{#1})}}
\newcommand{\es}[2]{{(\ref{#1}) and (\ref{#2})}}

\newcommand{\Ref}[1]{{Ref.~\cite{#1}}}

\newcommand{\equi}[1]{\stackrel{{#1}}{=}}

\newcommand{\ie}{{${ i.e., \ }$}}
\newcommand{\eg}{{${ e.g., \ }$}}

\newcommand{\cf}{{cf.\ }}

\newcommand{\aka}{{also known as }}
\newcommand{\wrt}{{with respect to }}
\newcommand{\wrtt}{{with respect to the }}

\newcommand{\lhs}{{left-hand side }}

\renewcommand{\~}{ \ }
\renewcommand{\=}{ \ = \ }
\newcommand{\eps}{\varepsilon^{}}

\newcommand{\p}{\!{}^{}}
\newcommand{\q}{{}^{}}

\newcommand{\Wedge}[1]{\bigwedge{}^{\!\! {#1}}}

\newcommand{\cL}{{\cal L}}

\newcommand{\rank}{{\rm rank}}

\newcommand{\id}{{\rm id}}

\newcommand{\twobyone}[2]{\left(\begin{array}{c}{#1} \cr
                                {#2} \end{array} \right)}

\newcommand{\twostack}[2]{\begin{array}{c} \lower.8ex\hbox{${#1}$}
                     \cr \raise.8ex\hbox{${#2}$} \end{array}}
\newcommand{\deder}[1]{{ 
 {\stackrel{\raise.2ex\hbox{$\leftarrow$}}{\delta^{r}}   } 
\over {   \delta {#1}}  }}
\newcommand{\dedel}[1]{{ 
 {\stackrel{\lower.3ex \hbox{$\rightarrow$}}{\delta^{\ell}}   }
 \over {   \delta {#1}}  }}
\newcommand{\papar}[1]{{ 
 {\stackrel{\raise.2ex\hbox{$\leftarrow$}}{\partial^{r}}   } 
\over {   \partial {#1}}  }}
\newcommand{\papal}[1]{{ 
 {\stackrel{\lower.3ex \hbox{$\rightarrow$}}{\partial^{\ell}}   }
 \over {   \partial {#1}}  }}
\newcommand{\rpa}[1]{{ 
 \stackrel{\raise.2ex\hbox{$\leftarrow$}}{\partial^{r}_{#1}}   }}
\newcommand{\lpa}[1]{{ 
 \stackrel{\lower.3ex\hbox{$\rightarrow$}}{\partial^{\ell}_{#1}}  }}

\newcommand{\proofbox}{\begin{flushright}{\hfill \ensuremath{\Box}}
\end{flushright}}

\newtheorem{theorem}{Theorem}[section]

\newtheorem{corollary}[theorem]{Corollary}

\newtheorem{definition}[theorem]{Definition}

\newtheorem{lemma}[theorem]{Lemma}

\newtheorem{proposition}[theorem]{Proposition}
\newtheorem{remark}[theorem]{Remark}

\begin{document}
\thispagestyle{empty}
\title{\Large{\bf Non-Decomposable Nambu Brackets}}
\author{{\sc Klaus~Bering}$^{a}$ \\~\\
$^{a}$Institute for Theoretical Physics \& Astrophysics\\
Masaryk University\\Kotl\'a\v{r}sk\'a 2\\CZ--611 37 Brno\\Czech Republic}
\maketitle
\vfill
\begin{abstract} 
It is well-known that the Fundamental Identity (FI) implies that Nambu brackets
are decomposable, \ie given by a determinantal formula. We find a weaker 
alternative to the FI that allows for non-decomposable Nambu brackets, but
still yields a Darboux-like Theorem via a Nambu-type generalization of
Weinstein's splitting principle for Poisson manifolds. 
\end{abstract}

\vfill

\begin{quote}
MSC number(s): 53D17; 53D99; 58A10; 70G10; 70G45; 70H50. 
\hrule width 5.cm \vskip 2.mm \noindent 
$^{a}${\small E--mail:~{\tt bering@physics.muni.cz}} \\
\end{quote}

\newpage

\section{Introduction}
\label{secintro}

\noi
Recall the definition of an almost Nambu-Poisson structure.

\begin{definition}
An {\bf almost $n$-Nambu-Poisson manifold} $(M;\pi)$ is a $d$-dimensional 
manifold $M$ with an $n$-multi-vector field 
\beq
\pi\=\frac{1}{n!}\pi^{i_{1}\ldots i_{n}}
\partial\q_{i_{1}}\wedge\dots\wedge\partial\q_{i_{n}}
\~\in\~\Gamma(\Wedge{n}TM)\~,
\eeq
with corresponding $n$-bracket 
$\{\cdot,\ldots,\cdot\}: [C^{\infty}(M)]^{\times n}\to C^{\infty}(M)$ defined as
\beq
\{f\q_{1},\ldots,f\q_{n}\}\= \pi(df_{1}\wedge\ldots\wedge df\q_{n})
\= \pi^{i_{1}\ldots i_{n}} 
\frac{\partial f\q_{1}}{\partial x^{i_{1}}} \ldots 
\frac{\partial f\q_{n}}{\partial x^{i_{n}}}\~, \qquad 
f\q_{1},\ldots,f\q_{n} \in C^{\infty}(M)\~,
\eeq
which is $\mathbb{R}$-multi-linear, totally skewsymmetric, and has the Poisson 
property (\ie Leibniz rule \wrt each entry). 
\end{definition}

\noi
The main question that we would like to discuss in this paper is:
``Which integrability conditions should one impose on the $n$-multi-vector 
field $\pi$?'' The case $n\!=\!1$ is just a vector field $\pi$, which has no 
non-trivial\footnote{
We will for simplicity not discuss the case where $M$ is a supermanifold, 
and/or where $\pi$ is Grassmann-odd. Recall that a Grassmann-odd vector field 
is not automatically in involution with itself.} 
integrability conditions. Moreover, the $n\!=\!1$ case is already manifestly 
decomposable --- in fact, it is what we call {\em decomposable Darboux}, \cf 
definition~\ref{defdecompdarboux}. {}For $n\!=\!2$, the bi-vector field $\pi$
should satisfy the Jacobi identity, and $(M;\pi)$ becomes a Poisson manifold. 
The sixty-four-thousand-dollar question is what should replace the Jacobi 
identity for $n\!\geq\!3$? Nambu himself left this question unanswered in his
seminal 1973 paper \cite{nambu73}. 

\noi
Twenty years later, in 1993, Takhtajan suggested to use the fundamental 
identity \e{fi1} as the missing integrability condition \cite{takhtajan93},
\cf Section~\ref{secfi}. We call such a structure a {\bf fundamental 
Nambu-Poisson structure.} Takhtajan also conjectured\footnote{Takhtajan likely
made the conjecture shortly after the publication of \Ref{takhtajan93}, see 
\Ref{dfst96} and Remark 6 in \Ref{takhtajan93}.} (and it 
was proven in 1996 by Gautheron \cite{gautheron96}) that the multi-vector field
$\pi$ then necessarily must be decomposable, \ie the $n$-bracket is given as a
determinant, \cf Theorem~\ref{fiimpliesdecomp}. This is surprisingly rigid
and in contrast to what happens in the $n\!=\!2$ Poisson case, where only the 
rank $2$ case is decomposable. Technically speaking, the culprit is the
fundamental algebraic identity \e{fai01}, \cf Section~\ref{secfai}, which is an
unavoidable consequence of the fundamental identity, \cf
Proposition~\ref{propfi}. More generally, a non-degenerately weighted 
fundamental algebraic identity \e{wfai01} necessarily implies pointwise 
decomposability, \cf Theorem~\ref{faiimpliesdecomp}, a result often attributed
to a 1996 paper \cite{alekguha96} by Alekseevsky and Guha, although it was
basically already known to Weitzenb\"ock \cite{weitzenbock23} in 1923.

\noi
One of the consequences of decomposability is as follows.  
Recall that the Cartesian product $M\q_{1}\!\times\! M\q_{2}$ of two Poisson 
manifolds $(M\q_{1};\pi\q_{1})$ and $(M\q_{2};\pi\q_{2})$ is again a Poisson 
manifold $(M\q_{1}\!\times\! M\q_{2};\pi\q_{1}\!+\!\pi\q_{2})$ by simply 
adding the two Poisson-bivectors $\pi\q_{i}\!\in\!\Gamma(\Wedge{2}TM\q_{i})$ 
together, $i\in\{1,2\}$. On the other hand, the Cartesian product 
$(M\q_{1}\!\times\! M\q_{2};\pi\q_{1}\!+\!\pi\q_{2})$ of two
$n$-Nambu-Poisson manifolds $(M\q_{1};\pi\q_{1})$ and $(M\q_{2};\pi\q_{2})$, 
where $\pi\q_{1}$ and $\pi\q_{2}$ are both $n$-multi-vector fields, that 
satisfy the fundamental identity, is {\em almost never} an $n$-Nambu manifold
itself for $n\!\geq\!3$, if one requires the fundamental identity to hold.  

\noi
One may ponder what decomposability means from a physics perspective? {}First 
a disclaimer. We have nothing new to say about the interesting and vast topic
of quantum Nambu brackets \cite{nambu73,dfst96}. Thus we are only discussing
classical physics, \ie the part of physics that does not dependent on Planck's
constant $\hbar$. Also we have nothing new to say about Nambu-type Hamiltonian
dynamics and equations of motion. Here we will only make a general comment
about kinematics. The decomposability issue does not affect Nambu structures
formulated on a world-volume $V$, as in membrane theory, \eg the recent 
Bagger-Lambert-Gustavsson (BLG) theory \cite{baglam06,gustavsson07,ai10}, 
because there the world-volume $V$ is of fixed low dimension, and one would
not be interested in forming Cartesian products of world volumes. Rather, the
issue arises in a field theoretic context with Nambu structures in the target
space. In the simplest Darboux case, one would formally have infinitely many
$n$-tuples of canonical field variables $\phi^{i}(x)$, $i\!=\!1, \ldots, n$, 
formally labeled by a continuous space-time index $x\!\in\!V$, \ie one is 
taking an infinite Cartesian product of Nambu structures.
 
\noi
Motivated by such considerations, we will abandon the fundamental identity
in this paper, and take another route. We are seeking a new definition of
$n$-Nambu-Poisson manifolds, that (as a consequence of yet-to-be-found 
conditions)
\begin{enumerate}
\item 
includes the decomposable case (where the $n$-bracket is given as a 
determinant, and where the fundamental identity is satisfied) as a special 
case;
\item
is stable under forming Cartesian products; 
\item
has a Darboux Theorem (in the form of a Weinstein splitting Theorem 
\cite{weinstein83}).
\end{enumerate}

\noi
Item 1 and 2 imply that one must allow $n$-multi-vector fields $\pi$
on Darboux form
\beq
\pi\=\sum_{m=1}^{r}\partial\q_{(m-1)n+1}\wedge\ldots\wedge\partial\q_{mn}\~,
\label{darbouxcoord00}
\eeq 
which are by definition non-decomposable when $r\!>\!1$, \cf 
Section~\ref{secdecomp}.  

\noi 
Another obstacles is related to the fact that not even a {\em pointwise} 
Darboux Theorem (as opposed to the usual neighborhood Darboux Theorem) holds 
for $n\!\geq\!3$. 

\noi
Perhaps the first idea is to replace the fundamental identity with a 
non-degenerately weighted generalized Poisson identity \e{wgpi01}, \cf 
Section~\ref{secwgps}. However, this seems not to be a feasible route for odd
$n\!\geq\!5$, and it is definitely excluded for $n\!=\!3$. In fact, we prove 
in the $n\!=\!3$ case, that a non-degenerately weighted generalized Poisson
identity \e{wgpi01} implies pointwise decomposability, \cf
Theorem~\ref{theoremwgpi03decomp}.

\noi
We have investigated various integrability and algebraic conditions in this 
paper. In the end, we choose to define a Nambu-Poisson structure as follows.

\begin{definition}
A {\bf Nambu-Poisson structure} is an almost Nambu-Poisson structure that
satisfied
\begin{enumerate}
\item 
the nested integrability property \e{fip01},
\item 
and the fundamental algebraic hyper-identity \e{fahi01}.
\end{enumerate}
\label{defintnambu}
\end{definition}

\noi
The algebraic condition $2$ in definition~\ref{defintnambu} help ensure a
pointwise Darboux Theorem, while condition $1$ is the actual integrability
condition. {}From these two assumptions we prove a Weinstein splitting
principle, \cf Theorem~\ref{weinsteinthm}. This is our main result.

\noi
{}Finally, we investigate in Appendix~\ref{secpresymp} if one may generalize
Moser's trick \cite{moser65} for symplectic $2$-forms to
$n$-pre-multi-symplectic forms with $n\!\geq\!3$. This seems not to be
generally possible, essentially because the flat map $\flat$ is almost never
surjective for $n\!\geq\!3$. However, for a limited result, see
Theorem~\ref{mosertheorem01}.

\section{Basic Formalism}

\noi
The {\bf sharp map} $\sharp:\Gamma(\Wedge{n-1}T^{*}M)\to\Gamma(TM)$ takes a
differential $n\!-\!1$ form 
\beq
\alpha\=\frac{1}{(n\!-\!1)!} \alpha\q_{i_{1}\ldots i_{n-1}}
dx^{i_{1}}\wedge \ldots \wedge dx^{i_{n-1}}\~\in\~\Gamma(\Wedge{n-1}T^{*}M)
\eeq
into a vector field
$\sharp(\alpha)^{j}\partial\q_{j}\!=\!\sharp(\alpha)\!=\!i\q_{\alpha}\pi$
with vector field components 
$\sharp(\alpha)^{j}=\alpha\q_{i_{1}\ldots i_{n-1}}\pi^{i_{1}\ldots i_{n-1}j}$.

\begin{definition}
The {\bf rank} of a multi-vector $\pi\q_{|\q_{p}}\!\in\!\Wedge{n}TM$ in a
point $p\!\in\!M$ is the dimension of the image of the sharp map, 
$\rank(\pi\q_{|\q_{p}}):=\dim({\rm Im}(\sharp\q_{|\q_{p}}))$. 
\end{definition}

\noi
The rank is lower semi-continuous as a function of the point $p\!\in\!M$.

\begin{definition}
A multi-vector $\pi\q_{|\q_{p}}\!\in\!\Wedge{n}TM$ is called 
{\bf non-degenerate} in a point $p\!\in\!M$ if the sharp map
$\sharp\q_{|\q_{p}}:\Wedge{n-1}T^{*}_{p}M \to T\p_{p}M$ is surjective, 
\ie if $\rank(\pi\q_{|\q_{p}})=d:=\dim(M)$. 
\end{definition}

\begin{definition}
An $n$-multi-vector field $\pi\!\in\!\Gamma(\Wedge{n}TM)$ is called {\bf 
invertible} if there exists an $n$-form $\omega\!\in\!\Gamma(\Wedge{n}T^{*}M)$
such that $J:=\sharp\circ\flat:TM\to TM$ is a pointwise invertible map,
where $\flat$ denotes the flat map, \cf Appendix~\ref{secpresymp}.
\end{definition}

\noi
An invertible $n$-multi-vector field is always non-degenerate.

\begin{definition}
A function $f\in C^{\infty}(M)$ is called a {\bf Casimir function} if 
$i\q_{df}\pi\!=\!0$. 
The {\bf center} $Z(M):=\{f\!\in\!C^{\infty}(M)\mid i\q_{df}\pi\!=\!0\}$
is the subalgebra of all Casimir functions. 
\end{definition}

\begin{definition}
A {\bf Hamiltonian vector field} is 
\beq
X\p_{\vec{f}} \~:=\~ \{f\q_{1},\ldots,f\q_{n-1},\~\cdot\~\}
\=\sharp(df_{1}\wedge\ldots\wedge df\q_{n-1})\~, \qquad 
f\q_{1},\ldots,f\q_{n-1} \in C^{\infty}(M)\~,
\eeq
and the $(n\!-\!1)$-tuple 
$\vec{f}:=(f\q_{1}, \ldots, f\q_{n-1})\in[C^{\infty}(M)]^{\times (n-1)}$ is
called a {\bf Hamiltonian}.
\end{definition}

\section{Pre-Combing the $n$-Bracket Locally}

\noi
In this Section we consider an arbitrary almost Nambu-Poisson structure 
$(M;\pi)$ without imposing any integrability conditions at all.

\begin{lemma}[Pre-Combing in a Neighborhood]
Let $\pi\!\in\!\Gamma(\Wedge{n}TM)$ be an $n$-multi-vector field with 
$n\!\geq\!2$. If the multi-vector $\pi\q_{|\q_{p}}\!\neq\!0$ is non-vanishing in
a point $p\!\in\!M$, then there exists a local coordinate system 
$(x^{1}, \ldots, x^{d})$ in a neighborhood $U$ of the point $p\!\in\!M$, such
that the Hamiltonian vector field
\beq
X\p_{(x^{1}, \ldots, x^{n-1})}\~\equiv\~\frac{\partial}{\partial x^{n}}\~,
\label{straigtennambu01}
\eeq
or equivalently, the corresponding $n$-bracket $\{\cdot,\ldots,\cdot\}$ 
fulfills  
\beq
\forall k\in\{1,\ldots, d\}\~:\~\~
 \{x^{1}, \ldots, x^{n-1}, x^{k} \} \~\equiv\~ \delta^{k}_{n}\~.
\label{straigtennambu02}
\eeq
\label{straigtennambulemma01}
\end{lemma}

\noi
{\sc Proof of Lemma~\ref{straigtennambulemma01}}:\~\~
One can choose local coordinates $(x^{1}, \ldots, x^{d})$ in a neighborhood $W$,
such that $\{x^{1},\ldots, x^{n}\}\q_{|\q_{p}}\neq\!0$, \ie such that 
$X\p_{(x^{1}, \ldots, x^{n-1})}\q_{|\q_{p}}\!\neq\!0$. One may always stratify
locally a non-vanishing vector field $X\p_{(x^{1}, \ldots, x^{n-1})}$ by choosing
new coordinates $(y^{1}, \ldots, y^{d})$ in a smaller neighborhood
$V\!\subseteq\!W$, such that 
$X\p_{(x^{1}, \ldots, x^{n-1})} \!=\!\partial / \partial y^{n}$. 
There must exist a subset of $n\!-\!1$ new coordinates 
$(y^{i_{1}}, \ldots, y^{i_{n-1}})$ with indices 
$i_{1}, \ldots, i_{n-1} \in \{1,\ldots, d\} $ such that the Jacobian
\beq
\det\left(\left.\frac{\partial x^{j}}{\partial y^{i_{k}}}
\right|_{p}\right)_{1\leq j,k\leq n-1} \~\neq \~0 
\eeq
is non-vanishing. Note that the $n\!-\!1$ indices 
$i_{1}, \ldots, i_{n-1} \neq n$ must all be different from index $n$, since
\beq
\forall j\in\{1,\ldots, n\!-\!1\}\~:\~\~  
\frac{\partial x^{j}}{\partial y^{n}}
\=\{x^{1}, \ldots, x^{n-1}, x^{j} \}\=0\~,
\eeq
because the $n$-bracket $\{\cdot,\ldots,\cdot\}$ is totally antisymmetric.
By relabeling the $y$-coordinates and perhaps shrinking to a smaller 
neighborhood $U\!\subseteq\!V$, one may assume that the Jacobian 
\beq
\det\left(\frac{\partial x^{j}}{\partial y^{k}}\right)_{1\leq j,k\leq n-1}
\~\neq \~0\~
\eeq
is non-vanishing in the whole neighborhood $U$. It is easy to check that the 
mixed coordinate system $(x^{1}, \ldots, x^{n-1},y^{n}, \ldots, y^{d})$ has
the sought-for properties \es{straigtennambu01}{straigtennambu02}. 
\proofbox

\begin{corollary}[Pre-Combing in a Point]
Let $\pi\!\in\!\Gamma(\Wedge{n}TM)$ be an $n$-multi-vector field with
$n\!\geq\!2$. If the multi-vector $\pi\q_{|\q_{p}}\!\neq\!0$ is non-vanishing 
in a point $p\!\in\!M$, then there exist local coordinates 
$(x^{1}, \ldots, x^{n},y^{n+1}, \ldots, y^{d})$ such that 
\beq
\{x^{1}, \ldots, x^{n} \}\q_{|\q_{p}} \=1\~, \label{straigtennambu03}
\eeq
and such that 
\beq
\forall j\in\{1,\ldots, n\}\forall k\in\{ n\!+\!1,\ldots, d\}\~:\~\~
 \{x^{1}, \ldots, \hat{x}^{j}, \ldots, x^{n}, y^{k} \}\q_{|\q_{p}} 
\= 0 \label{straigtennambu04}
\eeq
in the point $p\!\in\!M$.
\label{straigtennambulemma02}
\end{corollary}

\noi
{\sc Proof of Corollary~\ref{straigtennambulemma02}}:\~\~
By Lemma~\ref{straigtennambulemma01}, there exist local coordinates
$(x^{1}, \ldots, x^{n}, y^{n+1}, \ldots, y^{d})$ such that 
$\{x^{1}, \ldots, x^{n}\}\!=\!1$, and such that
\beq
\forall k\in\{ n\!+\!1,\ldots, d\}\~:\~\~
\{x^{1}, \ldots, x^{n-1},y^{k} \} \=0\~.
\eeq
Define new $y$-coordinates 
\beq
 y^{\prime k}\~:=\~y^{k} - \sum_{i=1}^{n}(-1)^{n-i}
\{x^{1}, \ldots, \hat{x}^{i}, \ldots, x^{n}, y^{k}\}(x^{j} - x^{j}_{|\q_{p}})
\eeq
for $k\!\in\!\{n\!+\!1,\ldots, d\}$. It is easy to check that the mixed 
coordinate system  
$(x^{1}, \ldots, x^{n},y^{\prime n+1}, \ldots, y^{\prime d})$ has the
sought-for property \e{straigtennambu04}. 
\proofbox

\section{Fundamental Identity}
\label{secfi}

\noi
The {\bf fundamental identity function}
$FI: [C^{\infty}(M)]^{\times (2n-1)}\to C^{\infty}(M)$ is defined by nested 
$n$-brackets as follows
\beq
{}FI(f\q_{1}, \ldots, f\q_{n-1}, g\q_{1},\ldots,g\q_{n})\~:=\~
X\p_{\vec{f}}\{g\q_{1},\ldots,g\q_{n}\}
-\sum_{i=1}^{n}\{g\q_{1},\ldots,g\q_{i-1},
X\p_{\vec{f}}[g\q_{i}],g\q_{i+1},\ldots,g\q_{n}\}\~.\label{fif}
\eeq

\begin{definition}
The {\bf fundamental identity} is \cite{takhtajan93}
\beq
{}FI(f\q_{1}, \ldots, f\q_{n-1}, g\q_{1},\ldots,g\q_{n})\=0\~,\label{fi0}
\eeq
or explicitly,
\beq
X\p_{\vec{f}}\{g\q_{1},\ldots,g\q_{n}\}\=
\sum_{i=1}^{n}\{g\q_{1},\ldots,g\q_{i-1},
X\p_{\vec{f}}[g\q_{i}],g\q_{i+1},\ldots,g\q_{n}\}\~, \label{fi1}
\eeq
or equivalently,
\beq
[X\p_{\vec{f}},X\p_{\vec{g}}]\=\sum_{i=1}^{n-1}X\p_{(g\q_{1},\ldots,g\q_{i-1},
X\p_{\vec{f}}[g\q_{i}],g\q_{i+1},\ldots,g\q_{n-1})}\~, \label{fi2}
\eeq
or equivalently, that Hamiltonian vector fields preserve the multi-vector 
field $\pi$,
\beq
{\cal L}\q_{X\p_{\vec{f}}}\pi\=0\~, \label{fi3}
\eeq
or equivalently,
\beq
X\p_{\vec{f}}\{g\q_{1},\ldots,g\q_{n}\}
\=\frac{1}{(n\!-\!1)!}\sum_{\sigma\in S\q_{n}}(-1)^{\sigma}
\{X\p_{\vec{f}}[g\q_{\sigma(1)}],g\q_{\sigma(2)},\ldots,g\q_{\sigma(n)}\}\~. 
\label{fi4}
\eeq
\label{fidef}
\end{definition}

\noi
The fundamental identity \e{fi1} was introduced in 1993 by Takhtajan
\cite{takhtajan93}.\footnote{
The fundamental identity in the $n\!=\!3$ case was considered in 1992 by Sahoo
and Valsakumar \cite{sahoovalsa92} under the name $5$-{\em point identity},
presumably because it has $2n\!-\!1\!=\!5$ entries. If one forgets about 
Leibniz rule, and think of $(C^{\infty}(M);\{\cdot,\ldots,\cdot\})$ as an 
infinite dimensional $n$-Lie algebra, the fundamental identity \e{fi1} was 
actually already introduced in 1985 by Filippov \cite{filippov85}.}

\section{Fundamental Algebraic Identity}
\label{secfai}

\begin{definition}
The {\bf fundamental algebraic identity} is
\beq
\sum_{i=1}^{n}\{h\q_{1},f\q_{1},\ldots,f\q_{n-2},g\q_{i}\}
\{g\q_{1},\ldots,g\q_{i-1},h\q_{2},g\q_{i+1},\ldots,g\q_{n}\}
\=-(h\q_{1} \leftrightarrow h\q_{2})\~, \label{fai01}
\eeq
or equivalently,
\beq
\sum_{\sigma\in S\q_{n}}(-1)^{\sigma}
\{h\q_{1},f\q_{1},\ldots,f\q_{n-2},g\q_{\sigma(1)}\}
\{g\q_{\sigma(2)},\ldots,g\q_{\sigma(n)},h\q_{2}\}
\=-(h\q_{1} \leftrightarrow h\q_{2})\~. \label{fai02}
\eeq
\label{faidef}
\end{definition}

\noi
The fundamental identity function \e{fif} satisfies Leibniz rule in each of 
its last $n$ entries $g\q_{1},\ldots,g\q_{n}$, but it does {\em not} satisfy 
Leibniz rule in each of its first $n\!-\!1$ entries 
$f\q_{1}, \ldots, f\q_{n-1}$ if $n\!\geq\!3$. In general, lack of Leibniz rule 
induces additional algebraic constraints. Concretely, 

\begin{proposition}
The fundamental identity \e{fi1} implies the fundamental algebraic identity 
\e{fai01}.
\label{propfi}
\end{proposition}

\noi
{\sc Proof of Proposition~\ref{propfi}:}\~\~Replace the entry
$f\q_{n-1}\!=\!h\q_{1}h\q_{2}$ in the fundamental identity \e{fi1} with a
product of functions.
\proofbox

\noi
The fundamental algebraic identity \e{fai01} is trivial for $n\!=\!2$.

\begin{remark}
The following tests are often useful in practice. 
\begin{itemize}
\item
To check if the fundamental algebraic identity \e{fai01} holds, it is enough to
test it locally, using only local coordinate functions $x^{1}, \ldots, x^{d}$ 
as entries. 
\item
If the fundamental algebraic identity \e{fai01} holds, to check 
if the fundamental identity \e{fi1} also holds, it is enough to test it
locally, using only local coordinate functions $x^{1}, \ldots, x^{d}$ as 
entries. 
\end{itemize}
\end{remark}

\noi
Similar practical tests exist for other identities below, although we will not 
always go into details.

\section{Fundamental Algebraic Hyper-Identity}

\begin{definition} The {\bf fundamental algebraic hyper-identity} is said to be 
satisfied if the fundamental algebraic identity 
\beq
\sum_{i=1}^{n}\{h\q_{1},f\q_{1},\ldots,f\q_{n-2},g\q_{i}\}
\{g\q_{1},\ldots,g\q_{i-1},h\q_{2},g\q_{i+1},\ldots,g\q_{n}\}
\=-(h\q_{1} \leftrightarrow h\q_{2}) \label{fahi01}
\eeq
holds for all $\mathbb{R}$-linearly dependent function tuples 
$(f\q_{1}, \ldots, f\q_{n-2}, g\q_{1}, \ldots, g\q_{n}, h\q_{1}, h\q_{2})$, \ie 
function tuples so that 
\beq
\exists (a\q_{1}, \ldots, a\q_{n-2}, b\q_{1}, \ldots, b\q_{n}, c\q_{1}, c\q_{2})
\in\mathbb{R}^{2n}\backslash\{\vec{0}\}:\~\~
\sum_{i=1}^{n-2} a\q_{i} f\q_{i} + \sum_{j=1}^{n} b\q_{j} g\q_{j} 
+ \sum_{k=1}^{2} c\q_{k} h\q_{k} \= 0\~.\label{lindep00}
\eeq
\end{definition}

\begin{remark} 
We mention the following practical test. 
\begin{itemize}
\item
To check if the fundamental algebraic  hyper-identity \e{fahi01} holds, 
it is enough to test it locally, using only local coordinate functions 
$x^{1}, \ldots, x^{d}$ as entries, where at least two entries are the 
same. 
\end{itemize}
\end{remark} 

\section{Weighted Fundamental Identity}

\begin{definition}
A {\bf weighted fundamental identity} is
\beq
X\p_{\vec{f}}\{g\q_{1},\ldots,g\q_{n}\}\=
\sum_{i=1}^{n}\lambda\q_{i}\{g\q_{1},\ldots,g\q_{i-1},
X\p_{\vec{f}}[g\q_{i}],g\q_{i+1},\ldots,g\q_{n}\}\~, \label{wfi1}
\eeq
with weight functions $\lambda\q_{i}\!\in\!C^{\infty}(M)$. 
\end{definition}

\noi
A weighted fundamental identity \e{wfi1} implies via symmetrization a
{\bf scaled fundamental identity} 
\beq
X\p_{\vec{f}}\{g\q_{1},\ldots,g\q_{n}\}\=\lambda
\sum_{i=1}^{n}\{g\q_{1},\ldots,g\q_{i-1},
X\p_{\vec{f}}[g\q_{i}],g\q_{i+1},\ldots,g\q_{n}\} \label{sfi1}
\eeq
with scale function
$\lambda=\frac{1}{n}\sum_{i=1}^{n}\lambda\q_{i}\in C^{\infty}(M)$.
A scaled fundamental identity \e{sfi1} implies an algebraic identity
\beq
(\lambda-1)X\p_{\vec{f}}[h\q_{1}] X\p_{\vec{g}}[h\q_{2}]
\=-(h\q_{1} \leftrightarrow h\q_{2})\~, \label{sfai01}  
\eeq
which can easily be seen by replacing the entry $g\q_{n}\!=\!h\q_{1}h\q_{2}$ 
in the scaled fundamental identity \e{sfi1} with a product of functions.
The algebraic identity \e{sfai01} implies that 
$(\lambda-1)\{f\q_{1},\ldots,f\q_{n}\}^{2}=0$, which immediately leads to 
the following alternatives:
\beq
\forall p\in M:\~\~ \lambda\q_{|\q_{p}}\=1 \~\~\vee\~\~ \pi\q_{|\q_{p}}\=0\~.  
\eeq
Conclusion: {\em There is nothing gained in terms of generality by
introducing weights $\lambda\q_{i}$ in the fundamental identity.}

\section{Weighted Fundamental Algebraic Identity}

\begin{definition}
A {\bf weighted fundamental algebraic identity} is 
\beq
\sum_{i=1}^{n}\lambda\q_{i}\{h\q_{1},f\q_{1},\ldots,f\q_{n-2},g\q_{i}\}
\{g\q_{1},\ldots,g\q_{i-1},h\q_{2},g\q_{i+1},\ldots,g\q_{n}\}
\=-(h\q_{1} \leftrightarrow h\q_{2})\~, \label{wfai01}
\eeq
with weight functions $\lambda\q_{i}\!\in\!C^{\infty}(M)$ that are
non-degenerate, \ie
\beq
\forall p\in M \exists i\in\{1,\ldots,n\}:\~\~\lambda\q_{i|\q_{p}}\~\neq\~0\~.
\label{lambdanondeg00}
\eeq 
\end{definition}

\begin{proposition}
A weighted fundamental identity \e{wfi1} implies a weighted fundamental 
algebraic identity \e{wfai01} with the same weights.
\label{propwfi}
\end{proposition}

\noi
{\sc Proof of Proposition~\ref{propwfi}:}\~\~Replace the entry
$f\q_{n-1}\!=\!h\q_{1}h\q_{2}$ in the weighted fundamental identity \e{wfi1}
with a product of functions.
\proofbox

\noi
The fundamental algebraic identity \e{fai01} is a special case of the
weighted fundamental algebraic identity \e{wfai01} with constant weights
$\lambda\q_{1}\!=\!\ldots\!=\!\lambda\q_{n}\!=\!1$. Conversely, the weighted
fundamental algebraic identity \e{wfai01} with non-vanishing average
$\frac{1}{n}\sum_{i=1}^{n}\lambda\q_{i}\!\neq\!0$ 
becomes a fundamental algebraic identity \e{fai01} via symmetrization.
In Corollary~\ref{threelittlepigs}, we prove that {\em there is nothing 
gained in terms of generality by introducing weights $\lambda\q_{i}$ in the 
fundamental algebraic identity.}

\begin{remark}[Normalization]
The non-degeneracy condition \e{lambdanondeg00} implies that locally in a
sufficiently small neighborhood $U\!\subseteq\!M$, it is possible to assume
that 
\beq
\lambda\q_{1|\q_{U}}\=1  \label{lambdanormal00}
\eeq
by relabeling and rescaling of the weighted fundamental algebraic identity
\e{wfai01}.
\label{remarknormlambda}
\end{remark}

\begin{definition} A {\bf weighted fundamental algebraic hyper-identity} is 
said to be satisfied if a weighted fundamental algebraic identity 
\beq
\sum_{i=1}^{n}\lambda\q_{i}\{h\q_{1},f\q_{1},\ldots,f\q_{n-2},g\q_{i}\}
\{g\q_{1},\ldots,g\q_{i-1},h\q_{2},g\q_{i+1},\ldots,g\q_{n}\}
\=-(h\q_{1} \leftrightarrow h\q_{2}) \label{wfahi01}
\eeq
holds for all $\mathbb{R}$-linearly dependent function tuples 
$(f\q_{1}, \ldots, f\q_{n-2}, g\q_{1}, \ldots, g\q_{n}, h\q_{1}, h\q_{2})$,
\cf \eq{lindep00}. 
\end{definition}

\section{Generalized Poisson Structure}

\begin{definition}
The {\bf generalized Poisson identity} \cite{app96,ai10} is
\beq
\sum_{\sigma\in S\q_{2n-1}}(-1)^{\sigma}\{f\q_{\sigma(1)},\ldots,f\q_{\sigma(n-1)},
\{f\q_{\sigma(n)},\ldots,f\q_{\sigma(2n-1)}\}\}\=0\~.
\label{gpi01}
\eeq 
The {\bf generalized algebraic Poisson identity} is
\beq
\sum_{\sigma\in S\q_{2n-2}}(-1)^{\sigma} 
\{h\q_{1},f\q_{\sigma(1)},\ldots,f\q_{\sigma(n-1)}\}
\{f\q_{\sigma(n)},\ldots,f\q_{\sigma(2n-2)},h\q_{2}\}
\=-(h\q_{1} \leftrightarrow h\q_{2})\~. \label{gapi01}
\eeq
\end{definition}

\begin{proposition}
The generalized Poisson identity \e{gpi01} implies the generalized algebraic 
Poisson identity \e{gapi01}.
\label{propgps}
\end{proposition}

\noi
{\sc Proof of Proposition~\ref{propgps}:}\~\~
Replace the entry $f\q_{2n-1}\!=\!h\q_{1}h\q_{2}$ in the generalized Poisson 
identity \e{gpi01} with a product of functions.
\proofbox

\begin{remark}
{}For even $n$, the generalized Poisson identity \e{gpi01} is equivalent to
involution 
\beq
(\pi,\pi)\q_{SN}\=0 \label{invo01}
\eeq
\wrtt Schouten-Nijenhuis antibracket 
$(\partial\q_{i},x^{j})\q_{SN}\!=\!\delta_{i}^{j}$.
{}For odd $n$, the involution condition \e{invo01} is trivially satisfied
because of the symmetry property of the Schouten-Nijenhuis antibracket.
\end{remark}

\begin{remark}
The fundamental identity \e{fi4} implies 
\bea
\lefteqn{
\sum_{\sigma\in S\q_{2n-2}}(-1)^{\sigma}
\{f\q_{\sigma(1)},\ldots,f\q_{\sigma(n-1)},
\{f\q_{\sigma(n)},\ldots,f\q_{\sigma(2n-2)},g\q_{1}\}\} } \cr
&=&n\sum_{\sigma\in S\q_{2n-2}}(-1)^{\sigma}
\{\{f\q_{\sigma(1)},\ldots,f\q_{\sigma(n)}\},
f\q_{\sigma(n+1)},\ldots,f\q_{\sigma(2n-2)},g\q_{1}\}\~, \label{mgpi01}
\eea
which, in turn, implies the generalized Poisson identity \e{gpi01}.
The identity \e{mgpi01} implies the algebraic identity
\beq
\sum_{\sigma\in S\q_{2n-3}}(-1)^{\sigma} 
\{h\q_{1},f\q_{\sigma(1)},\ldots,f\q_{\sigma(n-1)}\}
\{f\q_{\sigma(n)},\ldots,f\q_{\sigma(2n-3)},g\q_{1},h\q_{2}\}
\=-(h\q_{1} \leftrightarrow h\q_{2})\~, \label{mgapi01}
\eeq 
which can easily be seen by replacing the entry $f\q_{2n-2}\!=\!h\q_{1}h\q_{2}$ 
in the identity \e{mgpi01} with a product of functions. 
The algebraic identity \e{mgapi01} implies the generalized algebraic Poisson
identity \e{gapi01}, and for $n$ odd, the two algebraic identities
\es{gapi01}{mgapi01} are equivalent. {}Finally, consider the $180^{\circ}$ 
cyclic permutation 
\beq
\tau\~:=\~(n,\ldots,2n\!-\!2,1,\ldots,n\!-\!1)\~\in\~S\q_{2n-2}\~, 
\label{tau01}
\eeq
of permutation parity $(-1)^{\tau}=-(-1)^{n}$. The parity implies that the
generalized algebraic Poisson identity \e{gapi01} is trivially satisfied 
for even $n$.
\end{remark}

\begin{remark}
{}For completeness, let us also mention the algebraic identity 
\cite{michorvaisman99} 
\beq
(i\q_{\alpha}\pi) \wedge (i\q_{\beta}\pi) = 0\~, 
\qquad \alpha,\beta\in\Gamma(T^{*}M)\~, \label{gapi02} 
\eeq
or equivalently,
\beq
\sum_{\sigma\in S\q_{2n-2}}(-1)^{\sigma} 
\{h\q_{1},f\q_{\sigma(1)},\ldots,f\q_{\sigma(n-1)}\}
\{f\q_{\sigma(n)},\ldots,f\q_{\sigma(2n-2)},h\q_{2}\}
\=(-1)^{n}(h\q_{1} \leftrightarrow h\q_{2})\~, \label{gapi03}
\eeq
or equivalently,
\beq
\sum_{\sigma\in S\q_{2n-3}}(-1)^{\sigma} 
\{h\q_{1},f\q_{\sigma(1)},\ldots,f\q_{\sigma(n-1)}\}
\{f\q_{\sigma(n)},\ldots,f\q_{\sigma(2n-3)},g\q_{1},h\q_{2}\}
\=(-1)^{n}(h\q_{1} \leftrightarrow h\q_{2})\~, \label{gapi04}
\eeq 
which are equivalent to the algebraic identities \es{gapi01}{mgapi01} when $n$
is odd. 
\end{remark}

\section{Weighted Generalized Poisson Structures}
\label{secwgps}

\begin{definition}
A {\bf weighted generalized Poisson identity} is
\beq
\sum_{\sigma\in S\q_{2n-1}}(-1)^{\sigma}\mu(\sigma)
\{f\q_{\sigma(1)},\ldots,f\q_{\sigma(n-1)},
\{f\q_{\sigma(n)},\ldots,f\q_{\sigma(2n-1)}\}\}\=0\~,
\label{wgpi01}
\eeq 
with weight functions $\mu:M\times S\q_{2n-1}\to\mathbb{R}$ that are
non-degenerate, \ie 
\beq
\forall p\in M\~:\~\~ \mu\q_{|\q_{p}}\~\neq\~ 0\~,\label{cnondeg01}
\eeq
and $\mu\q_{|\q_{p}}:S\q_{2n-1}\to\mathbb{R}$ is symmetric in its first $n\!-\!1$
(and its last $n$) entries, respectively. Moreover, it is always assumed that
$\mu(\sigma)\!\in\!C^{\infty}(M)$ is a smooth function for each permutation
$\sigma\!\in\!S\q_{2n-1}$. 
\end{definition}

\noi 
The generalized Poisson structure \e{gpi01} is a special case of a weighted
generalized Poisson structure \e{wgpi01} with constant weights $\mu\!=\!1$.
Conversely, a weighted generalized Poisson structure \e{wgpi01} with
non-vanishing average
$\frac{1}{(2n\!-\!1)!}\sum_{\sigma\in S\q_{2n-1}}\mu(\sigma)\!\neq\!0$ becomes a
generalized Poisson structure \e{gpi01} by total antisymmetrization.

\begin{definition}
A {\bf weighted generalized algebraic Poisson identity} is
\beq
\sum_{\sigma\in S\q_{2n-2}}(-1)^{\sigma}\mu(\sigma)
\{h\q_{1},f\q_{\sigma(1)},\ldots,f\q_{\sigma(n-1)}\}
\{f\q_{\sigma(n)},\ldots,f\q_{\sigma(2n-2)},h\q_{2}\}
\=-(h\q_{1} \leftrightarrow h\q_{2})\~. \label{wgapi01}
\eeq 
with weight functions $\mu:M\times S\q_{2n-2}\to\mathbb{R}$ that are
non-degenerate, \ie 
\beq
\forall p\in M\~:\~\~ \mu\q_{|\q_{p}}\~\neq\~ 0\~,\label{cnondeg02}
\eeq
and $\mu\q_{|\q_{p}}:S\q_{2n-2}\to\mathbb{R}$ is symmetric in its first (and
last) $n\!-\!1$ entries, respectively. Moreover, it is always assumed that
$\mu(\sigma)\!\in\!C^{\infty}(M)$ is a smooth function for each permutation
$\sigma\!\in\!S\q_{2n-2}$. 
\end{definition}

\begin{remark}[Associated Weighted Generalized Algebraic Poisson Identities]
Consider some $k\!\in\!\{1,\ldots,2n\!-\!1\}$.
By replacing the entry $f\q_{k}\!=\!h\q_{1}h\q_{2}$ in the 
weighted generalized Poisson identity \e{wgpi01}, one derives
\beq
\sum_{\begin{array}{c}\sigma\in S\q_{2n-1} \cr 
{\scriptstyle\sigma(2n-1)=k }\end{array}} 
(-1)^{\sigma}\mu(\sigma)\{h\q_{1},f\q_{\sigma(1)},\ldots, f\q_{\sigma(n-1)}\}
\{f\q_{\sigma(n)},\ldots, f\q_{\sigma(2n-2)},h\q_{2}\}
\=-(h\q_{1} \leftrightarrow h\q_{2})\~, \label{wgapi02}
\eeq
which is of the form of a weighted generalized algebraic Poisson identity
\e{wgapi01}.
\label{obsremark01}
\end{remark}

\begin{remark}[Normalization]
The non-degeneracy conditions \e{cnondeg01} (or \e{cnondeg02}) imply that 
locally in a sufficiently small neighborhood $U\!\subseteq\!M$, it is possible
to assume that 
\beq
\mu\q_{|\q_{U \times \{\id\}}}\=1  \label{cnormal01}
\eeq
by relabeling and rescaling of the weighted identities \e{wgpi01} (or
\e{wgapi01}), respectively.
\label{remarknormc}
\end{remark}

\section{Integrability}

\begin{definition}
Given $2n\!-\!2$ functions $f\q_{1}, \ldots, f\q_{2n-2}\in C^{\infty}(M)$, 
the {\bf nested Hamiltonian distribution} is
\beq
\Delta\q_{2}(f\q_{1}, \ldots, f\q_{2n-2})
\~:=\~ {\rm span}\q_{C^{\infty}(M)}\left\{\left. 
X_{(X_{(f\q_{\sigma(1)},\ldots, f\q_{\sigma(n-1)})} [f\q_{\sigma(n)}],
f\q_{\sigma(n+1)}, \ldots, f\q_{\sigma(2n-2)})} 
\right| \sigma\in S\q_{2n-2}\right\}\~.\label{hamdistrib}  
\eeq
The {\bf nested integrability property} is
\beq
\forall f\q_{1},\ldots,f\q_{n-1},g\q_{1},\ldots,g\q_{n-1} \in C^{\infty}(M)\~:\~
[X\p_{\vec{f}},X\p_{\vec{g}}]\~\in\~\Delta\q_{2}(\vec{f},\vec{g})\~. \label{fip01}
\eeq
The {\bf Casimir integrability property} is
\beq
\forall f\q_{1},\ldots,f\q_{n} \in C^{\infty}(M)\~:\~ 
\{ f\q_{1},\ldots,f\q_{n}\} \in Z(M)\~\~ \Rightarrow \~\~
\left\{\begin{array}{l} {\rm The~}n{\rm~Hamiltonian~vector~fields} \\ 
X\p_{(\hat{f}\q_{1},f\q_{2},\ldots,f\q_{n})}, 
X\p_{(f\q_{1},\hat{f}\q_{2},f\q_{3},\ldots,f\q_{n})}, 
\ldots, 
X\p_{(f\q_{1},\ldots,f\q_{n-1},\hat{f}\q_{n})} \\
{\rm~are~in~involution}.
\end{array} \right. \label{fip02}
\eeq

\end{definition}

\begin{remark}
The fundamental identity \e{fi2} implies the nested integrability property 
\e{fip01}, which, in turn, implies the Casimir integrability property
\e{fip02}, and, with abuse of language, a weighted generalized Poisson identity 
\e{wgpi01}, where the weight functions $\mu(\sigma)$ may depend the input 
functions $f\q_{1}, \ldots, f\q_{2n-1}$.
\end{remark}

\noi
On the other hand, the generalized Poisson identity \e{gpi01} (or a weighted 
generalized Poisson identity \e{wgpi01}) does {\em not} necessarily have the 
nested integrability property \e{fip01} or the Casimir integrability property 
\e{fip02}. We can now prove a neighborhood version of 
Corollary~\ref{straigtennambulemma02}.

\begin{lemma}[Combing with the Casimir Integrability Property]
Let $\pi\!\in\!\Gamma(\Wedge{n}TM)$ be an $n$-multi-vector field that has the 
Casimir integrability property \e{fip02} with $n\!\geq\!2$.
If the multi-vector $\pi\q_{|\q_{p}}\!\neq\!0$ is non-vanishing in a point 
$p\!\in\!M$, then there exists a local coordinate system 
$(x^{1}, \ldots, x^{n},y^{n+1}, \ldots, y^{d})$ in a neighborhood $U$ of the
point $p\!\in\!M$ such that 
\beq
\{x^{1}, \ldots, x^{n} \}\=1\~, \label{straigtennambu05}
\eeq
and such that 
\beq
\forall j\in\{1,\ldots, n\}\forall k\in\{ n\!+\!1,\ldots, d\}\~:\~\~
\{x^{1}, \ldots, \hat{x}^{j}, \ldots, x^{n}, y^{k} \} 
\= 0 \label{straigtennambu06}
\eeq
in the whole neighborhood $U$.
\label{straigtennambulemma03}
\end{lemma}

\noi
{\sc Proof of Lemma~\ref{straigtennambulemma03}}:\~\~One may assume the
$\pi\q_{|\q_{p}}\!\neq\!0$. By Lemma~\ref{straigtennambulemma01}, there exists a
local coordinate system $(x^{1}, \ldots, x^{d})$ such that 
$\{x^{1}, \ldots, x^{n} \}\!=\!1$ in a neighborhood $V$. By the Casimir 
integrability property \e{fip02}, the $n$ Hamiltonian vector fields
\beq
X\p_{(\hat{x}^{1}, x^{2}, \ldots, x^{n})}, 
X\p_{(x^{1}, \hat{x}^{2}, x^{3}, \ldots, x^{n})}, \ldots, 
X\p_{(x^{1}, \ldots, x^{n-1}, \hat{x}^{n})}
\eeq
are in involution and linearly independent. By Frobenius Theorem, there exists
a coordinate system $(y^{1}, \ldots, y^{d})$ in a neighborhood
$U\!\subseteq\!V$ such that  
\beq
\forall j\in\{1,\ldots,n\}:\~\~
X\p_{(x^{1}, \ldots, \hat{x}^{j}, \ldots, x^{n})}
\=\frac{\partial}{\partial y^{j}}\~. 
\eeq
Since
\beq
\forall i,j \in\{1,\ldots, n\}\~:\~\~  
\frac{\partial x^{i}}{\partial y^{j}}
\=\{x^{1}, \ldots,\hat{x}^{j}, \ldots, x^{n}, x^{i} \}
\=(-1)^{n-j}\delta^{i}_{j}\~,
\eeq
the Jacobian 
\beq
\det\left(\frac{\partial x^{j}}{\partial y^{k}}\right)_{1\leq j,k\leq n}
\~\neq \~0\~
\eeq
is non-vanishing in the whole neighborhood $U$. The mixture
$(x^{1}, \ldots, x^{n},y^{n+1}, \ldots, y^{d})$ is therefore a coordinate 
system. It is easy to check that \eq{straigtennambu06} is satisfied.
\proofbox

\section{Decomposability and Darboux Coordinates}
\label{secdecomp}

\noi
\begin{definition}
An $n$-multi-vector field $\pi\!\in\!\Gamma(\Wedge{n}TM)$ is called 
{\bf (globally) decomposable} if there exist $n$ (globally defined) vector 
fields $X\p_{1}, \ldots, X\p_{n}\in\Gamma(TM)$ such that
\beq
\pi\=X\p_{1}\wedge\ldots\wedge X\p_{n}\~. 
\eeq
In other words, a {\bf decomposable $n$-bracket} is the same as a 
{\bf determinant $n$-bracket}
\beq
\{f\q_{1},\ldots,f\q_{n}\}\=\det(X\p_{i}[f\q_{j}])\~.\label{det01}
\eeq
\end{definition}

\begin{definition}
An $n$-multi-vector $\pi\q_{|\q_{p}}\!\in\!\Wedge{n}T\p_{p}M$ is said to be 
{\bf decomposable in a point $p\!\in\!M$}, if there exist $n$ vectors 
$X\p_{1|\q_{p}}, \ldots, X\p_{n|\q_{p}}\in T\p_{p}M$, such that
\beq
\pi\q_{|\q_{p}}\=X\p_{1|\q_{p}}\wedge\ldots\wedge X\p_{n|\q_{p}}\~.
\eeq
\end{definition}

\begin{definition}
A {\bf Darboux coordinate system}
$(x^{1}, \ldots, x^{nr},y^{nr+1}, \ldots, y^{d})$ in a local neighborhood $U$, 
where $r\!\in\!\{0,1,2,\ldots,[d/n]\}$, satisfies
\beq
\pi\q_{|\q_{U}}
\=\sum_{m=1}^{r}\partial\q_{(m-1)n+1}\wedge\ldots\wedge\partial\q_{mn}
\label{darbouxcoord01}
\eeq
in the whole neighborhood $U$.
\end{definition}

\noi
The rank, $\rank(\pi\q_{|\q_{U}})\!=\!nr$, of the $n$-multi vector field $\pi$
is then a multiplum of the order $n$, corresponding to that {\bf canonical
coordinates} $(x^{1}, \ldots, x^{nr})$ come in $n$-tuples. The $y$-coordinate
functions $y^{nr+1},\ldots, y^{d}$ are local Casimir functions in $U$.

\begin{definition}
A {\bf Weinstein split coordinate system} 
$(x^{1}, \ldots, x^{nr},y^{nr+1}, \ldots, y^{d})$ in a local neighborhood $U$
around a point $p\!\in\!M$, where $r\!\in\!\{0,1,2,\ldots,[d/n]\}$, satisfies
\beq
\pi\q_{|\q_{U}}
\=\sum_{m=1}^{r}\partial\q_{(m-1)n+1}\wedge\ldots\wedge\partial\q_{mn}
+\pi^{(y)}
\eeq
where the remainder $\pi^{(y)}\!\in\!\Gamma(\Wedge{n}TM\q_{|\q_{U}})$ is
independent of the $x$-coordinates $(x^{1}, \ldots, x^{nr})$ in the whole
neighborhood $U$, and where $\pi^{(y)}_{|\q_{p}}$ has vanishing rank, 
$\rank(\pi^{(y)}_{|\q_{p}})\!=\!0$, in the point $p\!\in\!M$.
\end{definition}

\noi
In particular, a Weinstein split coordinate patch 
$(U;\pi\q_{|\q_{U}})\!=\!(U^{(x)};\pi^{(x)})\!\times\!(U^{(y)};\pi^{(y)})$ is a 
product $U\!=\!U^{(x)}\!\times\!U^{(y)}$ of a Darboux patch $(U^{(x)};\pi^{(x)})$
and a patch $(U^{(y)};\pi^{(y)})$ with vanishing rank in at least one point.

\begin{definition}
An $n$-multi-vector field $\pi\!\in\!\Gamma(\Wedge{n}TM)$ is said to be 
{\bf decomposable Darboux}, if for all points $p\!\in\!M$ with
$\pi\q_{|\q_{p}}\neq 0$, there exist local coordinates $(x^{1},\ldots,x^{d})$ 
in a local neighborhood $U$ around $p\!\in\!M$ such that
\beq
\pi\q_{|\q_{U}}\=\partial\q_{1}\wedge\ldots\wedge\partial\q_{n}\~.
\label{decompdarboux}
\eeq
\label{defdecompdarboux}
\end{definition}

\section{Decomposable and Darboux Cases}

\begin{proposition}[Decomposable $\Rightarrow$ Fundamental Algebraic 
Identity]   
A multi-vector $\pi\q_{|\q_{p}}$ that is decomposable in a point $p\!\in\!M$ 
must satisfy the fundamental algebraic identity \e{fai02} in $p\!\in\!M$.
\label{propdecompimpliesfai}
\end{proposition}  

\noi
{\sc Proof of Proposition~\ref{propdecompimpliesfai}}:
This follows from the Schouten identity\footnote{Proof of the Schouten
identity~\e{si01}: One only has to consider non-zero contributions to 
\eq{si01}. In particular, one may assume that all indices take values inside 
$\{1,\ldots,n\}$ (where the Levi-Civita $\varepsilon$ symbol can be non-zero) 
rather than $\{1,\ldots,d\}$. If there are repetitions among 
$j\q_{1}, \ldots, j\q_{n}\in\{1,\ldots,n\}$, they must cancel out in the
alternating sum. Hence one may assume that $(j\q_{1}, \ldots, j\q_{n})$ is a
permutation of $(1,\ldots,n)$. It follows that $j\q_{\sigma(1)}\!=\!k\q_{2}$,
and hence that $k\q_{1}\!\neq\!k\q_{2}$. Moreover, there must exists 
$\ell\in\{2,\ldots,n\}$ such that $j\q_{\sigma(\ell)}\!=\!k\q_{1}$.
This contribution is canceled by a corresponding term in the second sum where 
$k\q_{1}\leftrightarrow k\q_{2}$ and $\sigma(1)\leftrightarrow \sigma(\ell)$ 
are both interchanged. \hfill$\Box$}
\beq
\sum_{\sigma\in S\q_{n}}(-1)^{\sigma} 
\~\varepsilon^{k\q_{1}i\q_{1}\ldots i\q_{n-2}j\q_{\sigma(1)}}\~
\varepsilon^{j\q_{\sigma(2)}\ldots j\q_{\sigma(n)}k\q_{2}}
\=-(k\q_{1} \leftrightarrow k\q_{2})\~. \label{si01}
\eeq
\proofbox

\begin{proposition}[Darboux $\Rightarrow$ Fundamental Algebraic 
Hyper-Identity]    
A multi-vector $\pi\q_{|\q_{p}}$ on Darboux form in a point $p\!\in\!M$ must
satisfy the fundamental algebraic hyper-identity \e{fahi01} in $p\!\in\!M$.
\label{propdarbouximpliesfahi}
\end{proposition}  

\noi
{\sc Proof of Proposition~\ref{propdarbouximpliesfahi}}:\~\~ One only
has to consider non-zero contributions to \eq{fahi01}. A non-zero contribution 
$\pi^{k\q_{1}i\q_{1}\ldots i\q_{n-2}j\q_{\sigma(1)}}\~
\pi^{j\q_{\sigma(2)}\ldots j\q_{\sigma(n)}k\q_{2}}$ must have indices 
$k\q_{1}, i\q_{1}, \ldots, i\q_{n-2}, j\q_{\sigma(1)}$ that belong to the same
canonical $n$-tuple, and similarly, the indices
$j\q_{\sigma(2)}, \ldots, j\q_{\sigma(n)}, k\q_{2}$
must belong to the same canonical $n$-tuple. So one may assume that all
the $2n$ indices fit within no more than 2 canonical $n$-tuples. If all
the indices belong to the same canonical $n$-tuple, the claim follows 
from the Schouten identity~\e{si01}. Now assume that $n$ indices belong to
one tuple and $n$ indices belong to a different tuple. By hyper-assumption, 
two indices must be the same. But this can only happen inside a tuple. But 
then the contribution vanish by skew-symmetry. 
\proofbox

\begin{proposition}[Decomposable Darboux $\Rightarrow$ Fundamental Identity]    
A decomposable Darboux multi-vector field $\pi\!\in\!\Gamma(\Wedge{n}T^{*}M)$
must satisfy the fundamental identity \e{fi1}. 
\label{propdecompdarbouximpliesfi}
\end{proposition}

\noi
{\sc Proof of Proposition~\ref{propdecompdarbouximpliesfi}}:\~\~This follows
from the pointwise observation (Proposition~\ref{propdecompimpliesfai}), and
the fact that the Levi-Civita $\eps$ symbol is $x$-independent.
\proofbox

\section{Weinstein Splitting Principle}

\noi
In this Section we prove converse statements to
Propositions~\ref{propdecompimpliesfai}, \ref{propdarbouximpliesfahi} and 
\ref{propdecompdarbouximpliesfi}.

\begin{lemma}[Combing with the Weighted Fundamental Algebraic
Hyper-Identity] Let $\pi\!\in\!\Gamma(\Wedge{n}TM)$ be an $n$-multi-vector 
field that satisfies a non-degenerately weighted fundamental algebraic 
hyper-identity \e{wfahi01} with $n\!\geq\!2$.
\begin{enumerate}
\item
If the multi-vector $\pi\q_{|\q_{p}}\!\neq\!0$ is non-vanishing in a point 
$p\!\in\!M$, then there exists a local coordinate system 
$(x^{1}, \ldots, x^{n},y^{n+1}, \ldots, y^{d})$ in a neighborhood $U$ of the
point $p\!\in\!M$ such that 
\beq
\{x^{1}, \ldots, x^{n}\}\q_{|\q_{p}}\=1\~, \label{straigtennambu07}
\eeq
and
\beq
\{x^{i_{1}}, \ldots, x^{i_{k}}, y^{i_{k+1}}, \ldots, y^{i_{n}} \}\q_{|\q_{p}} 
\= 0\~, \quad 
1\leq i_{1}<\ldots < i_{k} \leq n < i_{k+1} < \ldots < i_{n} \leq d\~, 
\quad 1\leq k<n\~,
\label{straigtennambu08}
\eeq
in the point $p\!\in\!M$.
\item
If furthermore the multi-vector $\pi\q_{|\q_{p}}$ satisfies 
a non-degenerately weighted fundamental algebraic identity \e{wfai01} or
a non-degenerately weighted generalized algebraic Poisson identity \e{wgapi01} 
in $p\!\in\!M$, then
\eq{straigtennambu08} holds for $k\!=\!0$ as well, \ie
\beq 
\{y^{i_{1}}, \ldots, y^{i_{n}} \}\q_{|\q_{p}}\!=\!0\~, \qquad 
n < i_{1} < \ldots < i_{n} \leq d\~. \label{straigtennambu09}
\eeq
In particular, the multi-vector 
$\pi\q_{|\q_{p}}=\partial\q_{1|\q_{p}}\wedge\ldots\wedge\partial\q_{n|\q_{p}}$ 
is decomposable in $p\!\in\!M$.   
\end{enumerate}
\label{faiimpliesdecompviggo}
\end{lemma}

\noi
{\sc Proof of part 1 of Lemma~\ref{faiimpliesdecompviggo}}:\~\~One may assume
the $\pi\q_{|\q_{p}}\!\neq\!0$. By Corollary~\ref{straigtennambulemma02}, there 
exist local coordinates $(x^{1}, \ldots, x^{n}, y^{n+1}, \ldots, y^{d})$ such 
that $\{x^{1}, \ldots, x^{n} \}\q_{|\q_{p}}\!=\!1$, and such that 
\beq
\{x^{i_{1}}, \ldots, x^{i_{n-1}}, y^{i_{n}} \}\q_{|\q_{p}} \= 0\~, \qquad 
1\leq i_{1}< \ldots < i_{n-1} \leq n < i_{n} \leq d\~, 
\label{straigtennambu10}
\eeq
which is just \eq{straigtennambu08} with $k\!=\!n\!-\!1$, \ie when there is
precisely one $y$-coordinate $y^{i_{n}}$ present on the \lhs of
\eq{straigtennambu08}. We would like to prove \eq{straigtennambu08} for 
{\em any} number $k$  of $x$-coordinates, where
$k\!\in\!\{1,\ldots, n\!-\!1\}$. So assume 
that $k\!\geq\!1$. Then there is at least one $x$-coordinate $x^{i_{1}}$ on the 
\lhs of \eq{straigtennambu08}. Since $k\!<\!n$, there must also be an 
$x$-coordinate $x^{\ell}$, $\ell\!\in\!\{1,\ldots,n\}$, that is {\em not}
present on the \lhs of \eq{straigtennambu08}. It is possible to normalize
the weight $\lambda\q_{1|\q_{p}}\!=\!1$ due to Remark~\ref{remarknormlambda}. 
Choose functions
$h\q_{1}\!=\!h\q_{2}\!=\!x^{i_{1}}$; $g\q_{1}\!=\!x^{\ell}$;
$f\q_{1}, \ldots, f\q_{n-2} \in \{x^{1}, \ldots, x^{n}\} 
\backslash \{x^{i_{1}},x^{\ell}\}$; and
$g\q_{2}, \ldots, g\q_{n} \in 
\{x^{i_{2}}, \ldots, x^{i_{k}}, y^{i_{k+1}}, \ldots, y^{i_{n}} \}$ in the weighted 
fundamental algebraic hyper-identity \e{wfahi01} . This proves 
\eq{straigtennambu08} for $k\!\in\!\{1,\ldots, n\!-\!1\}$.

\noi
{\sc Proof of part 2 of Lemma~\ref{faiimpliesdecompviggo}}:\~\~
{}Finally, consider the case $k\!=\!0$. Let us assume a weighted generalized
algebraic Poisson identity \e{wgapi01}. (The case of a weighted fundamental
algebraic identity \e{wfai01} is very similar.) Choose functions 
$f\q_{1}, \ldots, f\q_{2n-2}, h\q_{1}, h\q_{2} 
\in \{x^{1}, \ldots, x^{n}, y^{i_{1}}, \ldots, y^{i_{n}}\}$
in the weighted generalized algebraic Poisson identity \e{wgapi01}.
Make sure that the weight in front of the term  
$\{x^{1}, \ldots, x^{n}\}\q_{|\q_{p}}\{y^{i_{1}}, \ldots, y^{i_{n}} \}\q_{|\q_{p}}$
is non-vanishing.
\proofbox

\begin{theorem}[Non-Deg.\ Weighted Fund.\ Alg.\ Identity
$\Rightarrow$ Pointwise Decomposable \cite{weitzenbock23,alekguha96}]
If $n\!\geq\!3$, a non-degenerately weighted fundamental algebraic identity 
\e{wfai01} implies that the multi-vector $\pi\q_{|\q_{p}}$ is decomposable in 
the corresponding point $p\!\in\!M$.
\label{faiimpliesdecomp}
\end{theorem} 

\noi
{\sc Proof of Theorem~\ref{faiimpliesdecomp}}:\~\~A non-degenerately weighted
fundamental algebraic identity implies all the assumptions of
Lemma~\ref{faiimpliesdecompviggo}.
\proofbox

\begin{corollary}
Let $\pi\!\in\!\Gamma(\Wedge{n}TM)$ be an $n$-multi-vector with $n\!\geq\!3$.
The following conditions are equivalent.
\begin{enumerate}
\item
A non-degenerately weighted fundamental algebraic identity \e{wfai01} is
satisfied in $p\!\in\!M$.
\item
The multi-vector $\pi\q_{|\q_{p}}$ is decomposable.
\item
The fundamental algebraic identity \e{fai01} is satisfied in $p\!\in\!M$.
\end{enumerate}
\label{threelittlepigs}
\end{corollary}

\begin{theorem}[Weinstein Splitting Principle]
If $n\!\geq\!2$, the nested integrability property \e{fip01} and the 
fundamental algebraic hyper-identity \e{fahi01} imply that for every point 
$p\!\in\!M$ there exists a Weinstein split coordinate system in a local
neighborhood $U$ of $p\!\in\!M$.
\label{weinsteinthm}
\end{theorem}

\noi
{\sc Proof of Theorem~\ref{weinsteinthm}}:\~\~This proof essentially follows
Nakashima's proof of Theorem~\ref{fiimpliesdecomp}, \cf \Ref{nakanishi98} and
\Ref{vaisman99}, which use Weinstein splitting principle \cite{weinstein83}. 
One may assume the $\pi\q_{|\q_{p}}\!\neq\!0$.
By Lemma~\ref{straigtennambulemma03}, there exists a local coordinate system
$(x^{1}, \ldots, x^{d})$ such that $\{x^{1}, \ldots, x^{n} \}\!=\!1$, and such
that
\beq
\forall j\in\{1,\ldots, n\}\forall k\in\{ n\!+\!1,\ldots, d\}\~:\~\~
\{x^{1}, \ldots, \hat{x}^{j}, \ldots, x^{n}, y^{k} \} 
\= 0 \label{straigtennambu11}
\eeq
in the whole neighborhood $U$. Now continue the proof pointwise as in the 
proof of the first part of Lemma~\ref{faiimpliesdecompviggo} to establish
\eq{straigtennambu08} for each point $p\!\in\!U$.
Next use the nested integrability property \e{fip01} to the commutator
\beq
(-1)^{n-j}\frac{\partial}{\partial x^{j}}\{y^{i_{1}}, \ldots, y^{i_{n}}\}
\=[X\p_{(x^{1},\ldots, \hat{x}^{j}, \ldots, x^{n})},
X\p_{(y^{i_{1}},\ldots, \ldots, y^{i_{n-1}})}][y^{i_{n}}] \label{nestedflip}
\eeq
to deduce that the $n$-bracket $\{y^{i_{1}}, \ldots, y^{i_{n}}\}$ cannot 
depend on the coordinates $x^{j}$, $j\!\in\!\{1,\ldots,n\}$. Thus the 
manifold $M$ factorizes locally, and one may repeat the Weinstein splitting
argument as long as there remains non-zero rank left.  
\proofbox

\begin{theorem}[Fundamental identity $\Rightarrow$ Decomposable Darboux
Theorem \cite{gautheron96}]
If $n\!\geq\!3$, the fundamental identity \e{fi1} implies that $\pi$ is a
decomposable Darboux multi-vector field.
\label{fiimpliesdecomp}
\end{theorem}

\noi
{\sc Proof of Theorem~\ref{fiimpliesdecomp}}:\~\~This proof essentially
follows the proof of Theorem~\ref{weinsteinthm}, although now one has access
to the second part of Lemma~\ref{faiimpliesdecompviggo} as well, so 
the nested integrability argument \e{nestedflip} and the Weinstein splitting 
procedure becomes superfluous.
\proofbox

\begin{proposition} 
The determinant $n$-bracket \e{det01} satisfies the fundamental identity 
\e{fi1} if and only if for all points $p\!\in\!M$ with $\pi\q_{|\q_{p}}\neq 0$, 
the vector fields $X\p_{1}, \ldots, X\p_{n}\in\Gamma(TM)$ are in involution at 
the point $p\!\in\!M$.
\label{propdet}
\end{proposition}

\noi
{\sc Proof of the ``only if'' part of Proposition~\ref{propdet}}:\~\~
One may assume the $\pi\q_{|\q_{p}}\!\neq\!0$. One knows from
Theorem~\ref{fiimpliesdecomp} that the decomposable $n$-vector field 
$\pi\!=\!X\p_{1}\wedge\ldots\wedge X\p_{n}$ can be locally written as 
$\pi\q_{|\q_{U}}\!=\!\partial\q_{1}\wedge\dots\wedge\partial\q_{n}$, and one 
knows from $\pi\q_{|\q_{p}}\neq 0$ that $X\p_{1}, \ldots, X\p_{n}$ are pointwise
linearly independent in some neighborhood $U$ of the point $p\!\in\!M$. 
Thus the following two distributions
\beq
{\rm span}\q_{C^{\infty}(U)}\{X\p_{1}, \ldots, X\p_{n} \}
\= {\rm span}\q_{C^{\infty}(U)}\{\partial\q_{1}, \ldots, \partial\q_{n} \}
\eeq
are the same. Since the latter is in involution, so must the former be.
\proofbox

\section{The $n\!=\!3$ Case}

\noi
{}For $n\!\geq\!4$, the generalized algebraic Poisson identity \e{gapi01}
is different from the fundamental algebraic identity \e{fai01}.
However, in the $n\!=\!3$ case, the generalized algebraic Poisson identity 
\e{gapi01} is equivalent to the fundamental algebraic identity \e{fai01}.

\begin{remark}[Evading Algebraic Identity via Degeneracy] 
In this paper we are particularly interested in multi-vector fields, which are
not necessarily pointwise decomposable. Theorem~\ref{faiimpliesdecomp}
tells us to avoid imposing non-degenerately weighted fundamental algebraic
identities \e{wfai01}. Now suppose that one is given some weighted generalized
algebraic Poisson identity
\beq
\sum_{\sigma\in S\q_{4}}(-1)^{\sigma}\mu(\sigma)
\{h\q_{1},f\q_{\sigma(1)},f\q_{\sigma(2)}\}
\{f\q_{\sigma(3)},f\q_{\sigma(4)},h\q_{2}\}
\=-(h\q_{1} \leftrightarrow h\q_{2})\~. \label{wgapi03}
\eeq 
with $\twobyone{4}{2}\!=\!6$ weight functions $\mu(\sigma)$. It is easy to see
that it can always be rewritten into a weighted fundamental algebraic identity
\e{wfai01} (which one would like to avoid) with three weight functions
$\lambda\q_{1}, \lambda\q_{2}, \lambda\q_{3}$. The only hope to evade
decomposability is that the $\lambda\q_{i}$ weights might perhaps be 
degenerate (=zero), \cf \eq{lambdanondeg00}. In fact, $\lambda\q_{i}\!=\!0$ 
if and only if the $\mu(\sigma)$ weights in the weighted generalized algebraic
Poisson identity \e{wgapi03} satisfy
\beq
\forall\sigma\in\!S\q_{4}:\~\~\mu(\tau\circ\sigma)\=-\mu(\sigma)\~.
\eeq
Here $\tau\!:=\!(4,3,1,2)\!\in\!S\q_{2n-2=4}$ is the $180^{\circ}$ cyclic 
permutation of even permutation parity $(-1)^{\tau}=+1$.
\label{obsremark02}
\end{remark}

\begin{theorem}
In the $n\!=\!3$ case, an arbitrary non-degenerately weighted generalized
Poisson structure
\beq
\sum_{\sigma\in S\q_{5}}(-1)^{\sigma}\mu(\sigma)\{f\q_{\sigma(1)},f\q_{\sigma(2)},
\{f\q_{\sigma(3)},f\q_{\sigma(4)},f\q_{\sigma(5)}\}\}\=0
\label{wgpi03}
\eeq 
is always pointwise decomposable.
\label{theoremwgpi03decomp}
\end{theorem} 

\noi
{\sc Indirect proof of Theorem~\ref{theoremwgpi03decomp}}:\~\~
We cannot allow any non-degenerately weighted fundamental algebraic 
identities \e{wfai01}, \cf Theorem~\ref{faiimpliesdecomp}.
The weighted generalized Poisson identity \e{wgpi03} has 
$\twobyone{5}{2}\!=\!10$ weight functions $\mu(\sigma)$. {}As a shorthand 
let us from now on write $\mu(\sigma)$ as $\mu\q_{\sigma(1),\sigma(2)}$.
The weighted generalized Poisson identity \e{wgpi03} implies $2n\!-\!1\!=\!5$
associated weighted generalized algebraic Poisson identities of the form 
\e{wgapi03}, \cf Remark~\ref{obsremark01}. Because of Remark~\ref{obsremark02},
one must demand
\beq
k\=1\~:  \qquad  
\mu\q_{23}\=-\mu\q_{45} \~, \qquad  
\mu\q_{24}\=-\mu\q_{35} \~, \qquad  
\mu\q_{25}\=-\mu\q_{34} \~,
\eeq
\beq
k\=2\~:  \qquad  
\mu\q_{13}\=-\mu\q_{45} \~, \qquad  
\mu\q_{14}\=-\mu\q_{35} \~, \qquad  
\mu\q_{15}\=-\mu\q_{34} \~,
\eeq
\beq
k\=3\~:  \qquad  
\mu\q_{12}\=-\mu\q_{45} \~, \qquad  
\mu\q_{14}\=-\mu\q_{25} \~, \qquad  
\mu\q_{15}\=-\mu\q_{24} \~,
\eeq
\beq
k\=4\~:  \qquad  
\mu\q_{12}\=-\mu\q_{35} \~, \qquad  
\mu\q_{13}\=-\mu\q_{25} \~, \qquad  
\mu\q_{15}\=-\mu\q_{23} \~,
\eeq
\beq
k\=5\~:  \qquad  
\mu\q_{12}\=-\mu\q_{34} \~, \qquad  
\mu\q_{13}\=-\mu\q_{24} \~, \qquad 
\mu\q_{14}\=-\mu\q_{23} \~.
\eeq
It is not hard to check that this implies that the all coefficient
$\mu(\sigma)\!=\!0$ must vanish. This contradicts the non-degeneracy
\e{cnondeg01}. In other words, there is no identity \e{wgpi03} to start with.
\proofbox

\vspace{0.8cm}

\noi
{\sc Acknowledgement:}~The author thanks Igor A.~Batalin and Andrew Swann for
discussions. The work of K.B.\ is supported by the Grant agency of the Czech
republic under the grant P201/12/G028.

\appendix

\section{Pre-Multi-Symplectic Manifolds}
\label{secpresymp}

\noi
Let $M$ be a $d$-dimensional manifold, let $n\!\geq\!1$ be an integer, and let
\beq
Z^{n}(M):=\{\omega \in\Gamma(\Wedge{n}T^{*}M) \mid d\omega=0 \}
\eeq
denote the set of closed $n$-forms 
\beq
\omega\=\frac{1}{n!} \omega\q_{i_{1}\ldots i_{n}}
dx^{i_{1}}\wedge \ldots \wedge dx^{i_{n}}
\~\in\~\Gamma(\Wedge{n}T^{*}M)\~, \qquad d\omega\=0\~,
\eeq
on $M$. 

\begin{definition}
A closed $n$-form $\omega$ is called a {\bf pre-multi-symplectic $n$-form}, and
the pair $(M;\omega)$ is called an {\bf $n$-pre-multi-symplectic manifold}.
\end{definition}
 
\noi
The {\bf flat map} $\flat:\Gamma(TM)\to\Gamma(\Wedge{n-1}T^{*}M)$ takes 
a vector field $X\!=\!X^{j}\partial\q_{j}\!\in\!\Gamma(TM)$ into a 
differential $n\!-\!1$ form 
\beq
\frac{1}{(n\!-\!1)!}\flat(X)\q_{i_{1}\ldots i_{n-1}}
dx^{i_{1}}\wedge \ldots \wedge dx^{i_{n-1}}
\=\flat(X)\=i\q_{X}\omega\~\in\~\Gamma(\Wedge{n-1}T^{*}M)
\eeq
with components
$\flat(X)\q_{i_{1}\ldots i_{n-1}}\!=\!X^{j}\omega\q_{ji_{1}\ldots i_{n-1}}$.

\begin{definition}
The {\bf rank} of an $n$-form 
$\omega\q_{|\q_{p}}\!\in\!\Wedge{n}T^{*}M$ in a point $p\!\in\!M$ is the
dimension $d$ of the manifold $M$ minus the dimension the kernel of the 
flat map, $\rank(\omega\q_{|\q_{p}}):=d-\dim({\ker}(\flat\q_{|\q_{p}}))$. 
\end{definition}

\noi
Recall by Poincar\'e Lemma, there locally exists a {\bf pre-multi-symplectic 
potential $(n\!-\!1)$-form 
$\vartheta\!\in\!\Gamma(\Wedge{n-1}T^{*}M\q_{|\q{U}})$}, so that
$\omega\q_{|\q{U}}\!=\!d\vartheta$. 

\begin{definition}
A {\bf Darboux coordinate system}
$(x^{1}, \ldots, x^{nr},y^{nr+1}, \ldots, y^{d})$ in a local neighborhood $U$, 
where $r\!\in\!\{0,1,2,\ldots,[d/n]\}$, satisfies
\beq
\omega\q_{|\q_{U}}
\=\sum_{m=1}^{r}dx^{(m-1)n+1}\wedge\ldots\wedge dx^{mn}
\label{darbouxcoord02}
\eeq
in the whole neighborhood $U$.\footnote{Pandit and Gangal considered the 
$n\!=\!3$ case in \Ref{panditgandal96} and \Ref{panditgandal99}.
Beware that definitions vary from author to author.
In de Donder-Weyl theory (\aka covariant Hamiltonian field theory),
a {\bf Darboux coordinate system} in a neighborhood $U\!\subseteq\!M$ means 
that an $n$-pre-multi-symplectic manifold $M$ of dimension $d$ is locally 
isomorphic to a 
$(n\!-\!1)$-multi-cotangent bundle $U\cong\Wedge{n-1}T^{*}Q\q_{|\q{V}}$; where 
$Q$ is an $k$-dimensional position manifold; where $V\!\subseteq\!Q$ is
a neighborhood with position coordinates $(q^{1}, \ldots, q^{k})$; where the 
fibers in $(n\!-\!1)$-multi-cotangent bundle $\Wedge{n-1}T^{*}Q\q_{|\q{V}}$ 
have momentum coordinates $p\q_{\mu_{1}\ldots\mu_{n-1}}$ with 
$1\!\leq\!\mu_{1}\!<\!\ldots\!<\!\mu_{n-1}\!\leq\!k$; and where
the pre-multi-symplectic $n$-form is locally given as
\beq
\omega\q_{|\q_{U}}
=\frac{1}{(n\!-\!1)!}
\sum_{\mu_{1},\ldots,\mu_{n-1}=1}^{k}
dp\q_{\mu_{1}\ldots\mu_{n-1}}\wedge dq^{\mu_{1}}\wedge\ldots\wedge dq^{\mu_{n-1}}\~,
\label{darbouxcoord03}
\eeq
see \eg \Ref{martinlmp88} and \Ref{cidl99}. In particular, the dimensions must
in this case satisfy $d:=\dim(M)=k+\twobyone{k}{n-1}$.}
\end{definition}

\noi
The rank, $\rank(\omega\q_{|\q_{U}})\!=\!nr$, of the $n$-multi vector field 
$\pi$ is then a multiplum of the order $n$, corresponding to 
that {\bf canonical coordinates} $(x^{1}, \ldots, x^{nr})$ come in $n$-tuples.
The $y$-coordinate functions $y^{nr+1},\ldots, y^{d}$ are called local {\bf 
Casimir functions} in $U$.

\begin{definition} 
An $n$-form $\omega$ has a {\bf conformal vector field} $X$ with {\bf 
conformal weight function} $\lambda\!\in\!C^{\infty}(M)$ if
\beq
\cL\q_{X}\omega\=\lambda\omega\~.\label{cvf01}
\eeq 
\end{definition}

\begin{remark}
It follows from the proof of Poincar\'e Lemma that if a pre-multi-symplectic
$n$-form $\omega$ has Darboux coordinates in some neighborhood $U$, then there
exists a local conformal vector field $X\!\in\!\Gamma(TM\q_{|\q{U}})$ for
$\omega\q_{|\q{U}}$ with conformal weight $\lambda\!=\!1$, which can be made to
vanish $X\p_{|\q{p}}\!=\!0$ in any point $p\!\in\!U$.  
\end{remark}

\begin{definition}
An $n$-form $\omega\q_{|\q_{p}}\!\in\!\Wedge{n}T^{*}M$ is called 
{\bf non-degenerate} in a point $p\!\in\!M$ if the flat map
$\flat\q_{|\q_{p}}:T\p_{p}M\to\Wedge{n-1}T^{*}_{p}M$ is injective. 
\end{definition}

\noi
The rank of a non-degenerate $n$-form is just the dimension $d$ of the 
manifold.

\begin{definition}
An $n$-form $\omega\!\in\!\Gamma(\Wedge{n}T^{*}M)$ is called {\bf invertible} 
if there exists an $n$-multi-vector field 
$\pi\!\in\!\Gamma(\Wedge{n}TM)$ such that $\sharp\circ\flat:TM\to TM$ is a 
pointwise invertible map, \ie the map 
$J\q_{|\q_{p}}:=\sharp\q_{|\q_{p}}\circ\flat\q_{|\q_{p}}:T\p_{p}M\to T\p_{p}M$ is 
a bijection for all $p\!\in\!M$.
\end{definition}

\noi
An invertible $n$-form is always non-degenerate.

\begin{definition}
An {\bf $n$-multi-symplectic\footnote{
Beware that definitions may vary from author to author. {}For instance, relative
to our conventions, \Ref{baezhoffnungrogers08} shifts the order $n$ and calls 
a manifold with a non-degenerate closed $n$-form for an 
{\bf $(n\!-\!1)$-plectic manifold}. As another example, \Ref{awane92} calls a
manifold equipped with a certain kind of Lie-algebra-valued symplectic 
$2$-form for a {\bf $k$-symplectic manifold}.} 
manifold} $(M;\omega)$ is a $d$-dimensional manifold $M$ with an invertible 
closed $n$-form $\omega\!\in\!\Gamma(\Wedge{n}T^{*}M)$.
\end{definition}

\noi
We next salvage what we can from Moser's local trick for the $n\!=\!2$ case
\cite{moser65} when we consider general order $n\!\geq\!2$. 
Sadly, it isn't much, mainly
because the flat map $\flat:TM\to\Wedge{n-1}T^{*}M$ is {\em never} surjective 
for $n\!\geq\!3$ and $d\!\geq\!4$. 

\begin{theorem}[$n$th order version of Moser's local trick] 
Let there be given two non-degenerate pre-multi-symplectic $n$-forms 
$\omega\q_{0}, \omega\q_{1}\!\in\!\Gamma(\Wedge{n}T^{*}M)$ such that 
\begin{enumerate}
\item
their corresponding flat maps
$\flat\q_{|\q{\omega\q_{0}}}, \flat\q_{|\q{\omega\q_{1}}}$ 
have pointwise the same image, 
\beq
\Delta\~:=\~
{\rm Im}(\flat\q_{|\q{\omega\q_{0}}}) \= {\rm Im}(\flat\q_{|\q{\omega\q_{1}}})
\~\subseteq\~\Wedge{n-1}T^{*}M  \~;
\eeq
\item
they agree $\omega\q_{0|\q_{p}}\=\omega\q_{1|\q_{p}}$ in a point $p\!\in\!M$; 
\item
they have conformal vector fields 
$Y\p_{0},Y\p_{1}\in\Gamma(TM)$ with conformal weights $\lambda\!=\!1$;
\item
and the conformal vector fields $Y\p_{0},Y\p_{1}$ vanish in the point 
$p\!\in\!M$, 
\beq
Y\p_{0|\q{p}}\=0\=Y\p_{1|\q{p}}\~.
\eeq
\end{enumerate}
Then there exists two neighborhoods $U\p_{0}$ and $U\p_{1}$ of $p\!\in\!M$, and
a diffeomorphism $\Psi:U\p_{0}\to U\p_{1}$, with the point $p\!\in\!M$ as a
fixed point $\Psi(p)\!=\!p$, such that the pullback
$\Psi^{*}\omega\q_{1}\!=\!\omega\q_{0}$ in the neighborhood $U\p_{0}$.
\label{mosertheorem01}
\end{theorem}

\noi
{\sc Proof of Theorem~\ref{mosertheorem01}}:\~\~
One may define two pre-multi-symplectic potential $n\!-\!1$ forms
\beq
\vartheta\q_{i}\~:=\~\flat\q_{|\omega\q_{i}}(Y\p_{i})\=i\q_{Y\p_{i}}\omega\q_{i}
\~, \qquad  
\omega\q_{i}\=\cL\q_{Y\p_{i}}\omega\q_{i}\=[d,i\q_{Y\p_{i}}]\omega\q_{i}
\=d\vartheta\q_{i}\~, \qquad 
\vartheta\q_{i|\q{p}}\=0\~, \qquad i\~\in\~\{0,1\}\~.
\eeq
Next define convex linear combinations 
\beq
\omega\q_{t}\~:=\~t\omega\q_{0}+(1-t)\omega\q_{1}\~, \qquad
\vartheta\q_{t}\~:=\~t\vartheta\q_{0}+(1-t)\vartheta\q_{1}\~,  \qquad
\omega\q_{t}\=d\vartheta\q_{t}\~, \qquad 
t\~\in\~\mathbb{R}\~.
\eeq
Since $\omega\q_{t|\q_{p}}$ in the point $p\!\in\!M$ is independent of 
$t\!\in\mathbb{R}$, one may assume\footnote{
{}For instance, put $r(t)\!:=\!\sqrt{t^{2}+(1\!-\!t)^2}>0$ and define angle
$\varphi(t)\in]-\frac{\pi}{4},\frac{3\pi}{4}[$ via
$r(t)\exp(i\varphi(t))\!=\!t\!+\!i(1\!-\!t)$, where $t\!\in\!\mathbb{R}$.
By continuity, it must be possible to cover the line
$\{p\}\!\times\!\mathbb{R}\subseteq M\!\times\!\mathbb{R}$ with open 
box neighborhoods $U\p_{(k)} \times ]t^{\prime}_{(k)},t^{\prime\prime}_{(k)}[$
in which the map
$r(t)^{-1}\flat\q_{|\omega\q_{t}}
=\cos(\varphi(t))\flat\q_{|\omega\q_{0}}
\!+\!\sin(\varphi(t))\flat\q_{|\omega\q_{1}}$
is pointwise injective. Since $\exp(i\varphi(t))$ belongs to a 
compact set in $\mathbb{C}$, there exists a finite subcover that does the job.
Pick the set $U\!\subseteq\!M$ as a corresponding finite intersection, 
which must be open and include the point $p\!\in\!M$.}
(by perhaps restricting to a local neighborhood $U$ of the point $p\!\in\!M$) 
that $\flat\q_{|\omega\q_{t}}:
TM\q_{|\q_{U}}\to\Delta\q_{|\q_{U}}\subseteq\Wedge{n-1}T^{*}M\q_{|\q_{U}}$
is a pointwise injective map for all $t\!\in\!\mathbb{R}$.

\noi
Now a vector field $X\p_{t}$ is uniquely specified via
\beq
\flat\q_{|\omega\q_{t}}(X\p_{t})
\=\flat\q_{|\omega\q_{0}}(Y\p_{0})-\flat\q_{|\omega\q_{1}}(Y\p_{1})
\~\in\~\Delta\q_{|\q_{U}}\~,  \qquad t\~\in\~\mathbb{R}\~. \label{xeq01}
\eeq
The corresponding flow equation is
\beq
\frac{d\Psi\q_{t}(q)}{dt}\=X\p_{t|\q_{\Psi\q_{t}(q)}}\~, \qquad 
\Psi\q_{t=0}(q)\=q\~, \qquad q\~\in\~U \~. \label{ode01}
\eeq
Notice that $\vartheta\q_{0|\q_{p}}\!=\!0\!=\!\vartheta\q_{1|\q_{p}}$, so that
$X\p_{t|\q_{p}}\!=\!0$, and hence the constant solution $\Psi\q_{t}(p)\!=\!p$, 
$t\!\in\!\mathbb{R}$, is the unique solution in the point $p\!\in\!M$.
The ODE \e{ode01} has for each $q\!\in\!U$ a unique solution for
$t\!\in\![0,1]$ (by perhaps shrinking $U$ further). 
It remains to check that $\Psi\!:=\!\Psi\q_{t=1}$ is the sought-for 
diffeomorphism. One calculates 
\bea
\frac{d}{dt}\left(\Psi^{*}_{t}\omega\q_{t}\right)
&=&\Psi^{*}_{t}\left(\cL\q_{X\p_{t}}\omega\q_{t}+\frac{d}{dt}\omega\q_{t}\right)
\=\Psi^{*}_{t}\left([d,i\q_{X\p_{t}}]\omega\q_{t}+\omega\q_{1}-\omega\q_{0}\right)
\cr
&=&\Psi^{*}_{t}d\left(i\q_{X\p_{t}}\omega\q_{t}
+\vartheta\q_{1}-\vartheta\q_{0}\right)
\=\Psi^{*}_{t}d\left(\flat\q_{|\omega\q_{t}}(X\p_{t})
+\flat\q_{|\omega\q_{1}}(Y\p_{1})-\flat\q_{|\omega\q_{0}}(Y\p_{0})\right)
\!\equi{\e{xeq01}}\!0~.
\eea
So the $n$-form 
\beq
\Psi^{*}_{t}\omega\q_{t}
\=\left\{\begin{array}{ccl}\Psi^{*}_{0}\omega\q_{0}&=&\omega\q_{0} \cr
\Psi^{*}_{1}\omega\q_{1}&=&\Psi^{*}\omega\q_{1} \end{array} \right.
\eeq
does not depend on the parameter $t\!\in\![0,1]$.
\proofbox

\end{document}